\pdfoutput=1
\documentclass[letterpaper, paper,11pt]{AAS}	

\usepackage{bm}
\usepackage{amsmath}
\usepackage[colorlinks=true, pdfstartview=FitV, linkcolor=black, citecolor= black, urlcolor= black]{hyperref}
\usepackage{footnpag}			      	
\usepackage{amsfonts}
\usepackage{subcaption}
\usepackage{graphicx}
\usepackage[version=4]{mhchem}
\usepackage{siunitx}
\usepackage{longtable,tabularx}
\usepackage{amsthm}
\usepackage[numbers]{natbib}
\theoremstyle{definition}
\newtheorem{definition}{Definition}

\PaperNumber{21-402}

\begin{document}

\title{Precision Attitude Stabilization with Intermittent External Torque}

\author{S.P Arjun Ram\thanks{Graduate Research Assistant, Department of Aerospace Engineering and Engineering Mechanics, The University of Texas at Austin. Email: arjun.ram@utexas.edu} ~and
Maruthi R Akella\thanks{Ashley H. Priddy Centennial Professor, Department of Aerospace Engineering and Engineering Mechanics,The University of Texas at Austin. Email: makella@mail.utexas.edu, AIAA Associate Fellow, AAS Fellow}}

\maketitle{}

\begin{abstract}
The attitude stabilization of a micro-satellite employing a variable-amplitude cold-gas thruster which reflects as a time varying gain on the control input is considered. Existing literature uses a persistence filter based approach that typically leads to large control gains and torque inputs during specific time intervals corresponding to the ``on'' phase of the external actuation. This work aims at reducing the transient spikes placed upon the torque commands by the judicious introduction of an additional time varying scaling signal as part of the control law. The time-update mechanism for the new scaling factor and overall closed-loop stability are established through a Lyapunov-like analysis. Numerical simulations highlight the various features of this new control algorithm for spacecraft attitude stabilization subject to torque intermittence. .
\end{abstract}

\section{Introduction}

The problem of attitude stabilization of a microsatellite employing a variable-amplitude cold-gas thruster so as to ensure zero torque commands during thruster rise and fall times is considered. Thrusters have been a popular choice for microsatellite attitude control systems owing to their high power-to-weight ratio and lesser number of moving parts. Thruster applications to spacecraft attitude stabilization have been widely studied in the past [\citenum{sidi}][\citenum{bryson}][\citenum{Stanton}]. There have been numerous developments in the design of variable amplitude thrusting without excessive loss of efficiency of operation. However, conventional gas-based thrusters have nonzero rise and fall times [\citenum{wertz}] and it is important to ensure that attitude control torque commands be implemented during the maximum thrust phase, since the thrust provided during the rise and fall phases is uncertain. Past literature focuses on this problem and provides a persistence filter based solution [\citenum{LettersSrikant}][\citenum{JGCDSrikant}] which however, has the potential disadvantage of commanding large control inputs that could be too high for the actuators to handle during the ``on'' phase of the thruster cycle. This work aims at modifying the persistence filter approach by replacing the high gain terms with a dynamically adjusted scaling factor thereby helping guard against potential control saturations.

To model this controller, an artificial control prescaling $g(t)$ is defined, typically chosen to be a periodically modulated step function representing the thruster on and off schedule, which can possibly be zero over several windows of time. For practical uses, $g(t)$ should resemble a modulated step function, that is zero during the thruster off or rise and fall times, and nonzero when the thrusters are in operation. Such problems where the control is scaled by a periodically singular gain is discussed by Loria et al. [\citenum{Loria}] and the persistently exciting case by Jiang et al. [\citenum{Jiang}]. A classical definition of persistence of excitation is given as follows [\citenum{Sastry}].

\begin{definition}[Persistent Excitation] 
A function $g(\cdot)$ is said to be persistently exciting (PE) if there exist finite positive constants $\mu_1$, $\mu_2$ and $T$, such that:
\begin{equation}
\mu_1 \leq \int_{t}^{t+T} g^2(\tau) d\tau \leq \mu_2 \quad \forall \quad t \geq 0
\end{equation}
\label{PEDef}
\end{definition}

An important feature of persistently exciting signals is that they can be singular (i.e. identically zero) at specific instants of time or possibly over extended time intervals (of finite duration). As a result, standard feedback linearization techniques do not apply and controller design for systems with scaled PE signals is challenging and non-trivial. In order to further motivate this discussion, consider a prototype linear dynamical system with PE control gains described by:

\begin{equation}
\bm{\dot{x}} = \bm{Ax} + g(t)\bm{Bu}
\label{linearEq}
\end{equation}
where $\bm{A}\in \mathbb{R}^{n\times n}$, $\bm{B} \in \mathbb{R}^{n\times m}$, pair $(\bm{A},\bm{B})$ controllable, $\bm{x}(t) \in \mathbb{R}^n$ is the system state, and $u(t) \in \mathbb{R}^m$ is the control input. The control input gain $g( \cdot ) : \mathbb{R}^{\geq 0} \rightarrow \mathbb{R}$ is a scalar PE signal following Definition~\ref{PEDef}. 

Lyapunov methods for linear time-varying systems with PE are explored by Maghenem and Loria [\citenum{Maghenem}]. Chitour et al. [\citenum{Chitour1}], [\citenum{Chitour2}] proved that when $\bm{A}$ is neutrally stable and $g(t) : \mathbb{R}^{\geq 0} \rightarrow [0, 1]$, system given in Eq. \ref{linearEq} can be stabilized by standard linear full-state feedback controllers. Furthermore, Ref. [\citenum{Mazanti1}] shows that it is possible to achieve arbitrary rate of convergence (exponential) for the two-dimensional special case in which the input $u$ is a scalar signal.Ref. [\citenum{Mazanti2}] analyzed the stability of system Eq. \ref{linearEq} under a time-delay feedback control law.

In this paper, we use Euler parameters (quaternions) to represent the attitude kinematics, which follow the Euler rotational dynamics, with the input torque scaled by the control gain $g(t)$. We emphasize that this $g(t)$ signal does not need to be periodic with time. This formulation is similar to that of magnetically actuated spacecraft, where the control scaling turns out to be a matrix with both state and time dependent terms [\citenum{Lovera1}][\citenum{Lovera2}]. Since $g(t)$ can be singular, we cannot simply divide it out of the dynamics involving the torque and need to come up with better approaches. The stabilization problem considered in this work is tackled using a persistence filter which was first developed by Srikant and Akella [\citenum{LettersSrikant}] for linear single-input systems. This was then extended for the filter state to have both state and time dependence [\citenum{JGCDSrikant}] and the controller was shown to guarantee exponential convergence of the angular velocity and the vector part of the attitude quaternion to zero for attitude stabilization problems. The backstepping control input design however, involved the reciprocal of the filter state which while bounded, can reach large values leading to spikes in the commanded torque inputs in the transient. The same approach but for multi-input linear systems with inputs that are scaled by a time-varying, singular gain matrix was explored by Srikant and Akella [\citenum{Srikant2015}] and a specific version of the problem by Dong and Akella [\citenum{Hongyang}].

This work aims at modifying the persistence filter approach to avoid these spikes and provide lower control inputs by replacing the high gain terms with a dynamically adjusted scaling factor, albeit at the cost of sacrificing exponential stability guarantees and replacing them with asymptotic stability. The need for a different control law and the tradeoffs will be elaborated further in the Motivation section of this paper.

Throughout this paper, boldface variables are used to denote vector quantities, and uppercase letters are used to denote matrices. $\|\cdot\|$ denotes the Euclidean norm for vectors in $\mathbb{R}^3$ and for matrices, it represents the Euclidean induced norm. The time argument for functions is left out unless necessary. The remainder of this paper is organized as follows: the system dynamics are described in the next section, followed by a section motivating this work compared to previous literature. The controller design and Lyapunov like analysis then follows. Simulations demonstrating the controller implementation are presented before we conclude the paper.
\section{Kinematics and Dynamics}

We consider the attitude stabilization problem with thruster on-off scheduling for spacecraft. Several attitude parameterizations are available in the literature [\citenum{Junkins}] to represent the attitude kinematics. In this work, the attitude kinematics are formulated using the four-dimensional unit-norm constrained Euler paramters (quaternions), $\bm{\beta} = [\beta_0, \bm{\beta_v}]^T$. The quaternion kinematic differential equation is given by: [\citenum{SchaubJunkins}]
\begin{equation}
\begin{bmatrix}
\dot{\beta}_0 \\ \bm{\dot{\beta}_v}
\end{bmatrix}
= \frac{1}{2}
\begin{bmatrix}
-\bm{\beta_t}^T \\
\beta_0\mathbb{I} + S(\bm{\beta_v})
\end{bmatrix}
\bm{\omega}
\label{quat}
\end{equation}
where $\bm{\beta} \in \mathbb{S}^3$, $\bm{\omega} \in \mathbb{R}^3$ is the spacecraft angular rate expressed in a body-fixed frame, $\mathbb{I}$ is the identity matrix in $\mathbb{R}^{3\times 3}$, and $S()$ represents the skew-symmetric matrix representing vector cross-product in 3-dimensions. The spacecraft dynamics represented in the body frame, along with a known thruster scheduling $g(t)$, are given by
\begin{equation}
\bm{J\dot{\omega}} = -\bm{S}(\bm{\omega})\bm{J\omega} + g(t)\bm{u}(t)
\label{dynamics}
\end{equation}
where $\bm{J}$ is the symmetric and positive definite spacecraft inertia tensor, $g(t)\bm{u}(t) \in \mathbb{R}^3$ is the actual commanded thruster torque, $\bm{u}(t) \in \mathbb{R}^3$ is the designated torque, and $g(t)$ is the scalar control gain function of our design. 

The goal of this paper is to formulate a full-state feedback controller which commands the torque input $u(t)$ to stabilize to the origin the angular velocity and the vector part of the attitude quaternion for the dynamics represented in Eq.~(\ref{quat}) and Eq.~(\ref{dynamics}), with a known pre-designed scheduling function $g(t)$.
\section{Motivation and Persistence Filter Design}
\label{Motivation}
\begin{figure}
\centering
\includegraphics[scale=0.5]{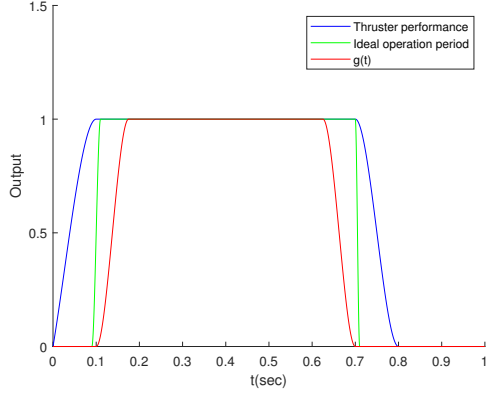}
\caption{Designing $g(t)$}
\label{gt}
\end{figure}

The persistence filter approach was proposed to handle the possibly singular control gains [\citenum{LettersSrikant}][\citenum{JGCDSrikant}]. However to ensure a positive lower bound on the filter state, some additional assumptions are placed upon the scaling function $g(t)$. To guarantee system controllability, $g(t)$ is assumed to be persistently exciting and the scalar part of the quaternion $\beta_0 = -1$ only during isolated instants of time. Moreover, for our controller, $g(t)$ will be required to be at least $\mathbb{C}^1$, which can be generated by approximating the actual discontinuous signal to $\mathbb{C}^1$. Fig. \ref{gt} shows a typical thruster performance plot and the ideal period during which input should be provided. The function $g(t)$ is chosen to resemble the step function while being $\mathbb{C}^1$ and having a non-infinite slope at the rise and fall times.

The persistent filter state $R(t) \in \mathbb{R}$ is defined to have the following dynamics [\citenum{JGCDSrikant}]:
\begin{equation}
\dot{R} = -\lambda R + g^2(1 + \beta_0)
\end{equation}
with the initial condition $R(0)>0$ for any $\lambda > 0$. It has been shown [\citenum{JGCDSrikant}] that under the PE assumptions placed on $g(t)$, $\exists R_{\rm min}>0$ such that $\forall t \geq 0$,
\begin{equation}
0 < R_{\rm min} \leq R(t) \leq \frac{2\|g\|_\infty^2}{\lambda} 
\end{equation}
where $\|\cdot \|_\infty$ denotes the infinity norm of the signal.

It was further shown in [\citenum{JGCDSrikant}] that the control input given by 
\begin{equation}
u = -\frac{2g}{R}S(\bm{\beta_v})\bm{J\omega} - \frac{4\dot{g}}{R}\bm{J\beta_v} + \frac{2g\dot{R}}{R^2}\bm{\beta_v} - \frac{2g}{R}\bm{J\dot{\beta}_v} - \frac{g}{2R}(1 + \beta_0)\bm{J\Omega}
\label{OldControl}
\end{equation}
with the augmented state
\begin{equation}
\bm{\Omega} = \bm{\omega} + \frac{2g^2}{R}\bm{\beta_v}
\end{equation}
guarantees the exponential convergence of $\lim_{t\rightarrow\infty}\bm{\beta}_v = 0$ and $\lim_{t\rightarrow\infty}\bm{\omega} = 0$, provided the choice of the filter bandwidth parameter $\lambda$ satisfies
\begin{equation}
\lambda > \max \left\lbrace\frac{1}{2J_{\rm min}},\frac{1}{2}\right\rbrace
\label{Lambda}
\end{equation}

\begin{figure}
\centering
\includegraphics[scale=0.5]{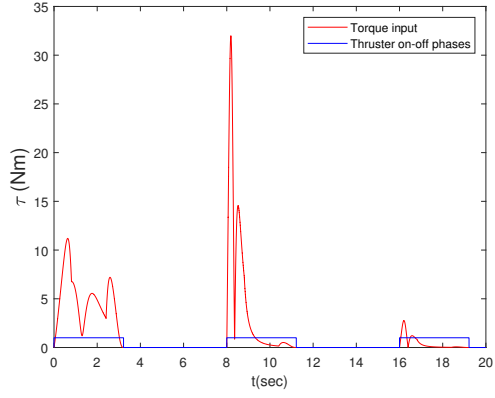}
\caption{Typical plot of input torque from [\citenum{JGCDSrikant}] superimposed with thruster schedule}
\label{Oldplot}
\end{figure}

The output state $R$ of the persistence filter updates as a function of $g(t)$ and has an enforced positive lower bound $R_{min}$. As the length of the inactive periods of the thruster increases, $R_{min}$ can go to arbitrarily low values which cannot be pre-determined. Thus as is evident from Eq.\ref{OldControl}, the reciprocal of $R$ can reach large values in the transient which leads to large values of the control input. A typical plot of the input torque using the controller from Eq.\ref{OldControl} superimposed with the thruster on-off schedule is shown in Fig.\ref{Oldplot}, which shows the spikes in the transient every time the thruster turns on. Since large control inputs are not practical, this work aims at replacing the $1/R$ terms in the control input with a newly designed dynamic (i.e. time-varying) scaling term $\hat{k}$, which updates as a function of the states and $g(t)$, while also using the persistence filter state $R$. The update mechanism for $\hat{k}$ also involves a learning rate parameter $\gamma > 0$ that controls the rate of change of $\hat{k}$ and acts as an extra tuning knob to choose how aggressive we want the control response to be, a feature that was not available before.
\section{Controller Design}

The proposed control input is given by:
\begin{equation}
\bm{u} = -2g\hat{k}S(\bm{\beta_v})\bm{J\omega} - 4\dot{g}\hat{k}\bm{J\beta_v} - 2g\dot{\hat{k}}\bm{J\beta_v} - 2g\hat{k}\bm{J\dot{\beta}_v} - \frac{g}{2}\hat{k}(1 + \beta_0)\bm{J\Omega}
\label{ControlInput}
\end{equation}
for the augmented state
\begin{equation}
\bm{\Omega} = \bm{\omega} + 2g^2\hat{k}\bm{\beta_v}
\end{equation}
which has the dynamics
\begin{equation}
\bm{J\dot{\Omega}} = -S(\bm{\omega})\bm{J\omega} - \frac{g^2}{2}\hat{k}(1+\beta_0)\bm{J\Omega}
\end{equation}
for the control input in Eq.\ref{ControlInput}.

We define an error term
\begin{equation}
\tilde{k} = \hat{k} - k^\ast
\end{equation}
where $k^\ast$ is a positive constant that will be defined later.

We add a positive term to the storage function from [\citenum{JGCDSrikant}]:
\begin{equation}
V = 2(1 - \beta_0)R + R\bm{\Omega^T J\Omega} + \frac{1}{2\gamma}\tilde{k}^2
\end{equation}
Taking the time derivative of which leads to
\begin{align}
\dot{V} &= -\lambda R\bm{\Omega^TJ\Omega} - 2\lambda (1 - \beta_0)R + R\bm{\beta_v^T\omega} + 2g^2\|\bm{\beta_v}\|^2 \nonumber \\ &+ (1 - R\hat{k})g^2(1+\beta_0)\bm{\Omega^T J\Omega} + \frac{1}{\gamma}\tilde{k}\dot{\hat{k}}
\end{align}

If we use the lower bound on the persistence filter state $R_{\rm min}$, the $(1 - R\hat{k})$ term can be replaced by the higher value $(R/R_{\rm min} - R\hat{k})$ term and multiply the non-negative term $2g^2\|\bm{\beta_v}\|$ by  the factor $R/R_{\rm min} \ge 1$ , leaving us with
\begin{align}
\dot{V} &\leq -\lambda R\bm{\Omega^TJ\Omega} - 2\lambda (1 - \beta_0)R + R\bm{\beta_v^T\omega} + 2g^2\frac{R}{R_{\rm min}}\|\bm{\beta_v}\|^2 \nonumber \\ &\quad + (\frac{R}{R_{\rm min}} - R\hat{k})g^2(1+\beta_0)\bm{\Omega^T J\Omega} + \frac{1}{\gamma}\tilde{k}\dot{\hat{k}}
\end{align}
If we fix the value of $k^\ast = 1/R_{\rm min}$, this leads to:
\begin{align}
\dot{V} &\leq -\lambda R\bm{\Omega^TJ\Omega} - 2\lambda (1 - \beta_0)R + R\bm{\beta_v^T\Omega} \nonumber \\ &\quad + \frac{\tilde{k}}{\gamma}\left(\dot{\hat{k}} - \gamma Rg^2(1+\beta_0)\bm{\Omega^TJ\Omega} - 2g^2R\gamma\|\bm{\beta_v}\|^2\right)
\end{align}
We now choose the update law for the scaling factor $\hat{k}$ to be
\begin{equation}
\dot{\hat{k}} = \gamma Rg^2(1+\beta_0)\bm{\Omega^T J\Omega} - 2\gamma g^2R\|\bm{\beta_v}\|
\end{equation}

which provides 
\begin{equation}
\dot{V} \leq \left( \frac{-\lambda J_{\rm min}}{J_{\rm max}} - \frac{1}{2J_{\rm max}}\right) RJ_{\rm max}\|\bm{\Omega}\|^2 - 2\left(\lambda - \frac{1+\beta_0}{4}\right) R(1-\beta_0)
\end{equation}
where $J_{\rm max}$ and $J_{\rm min}$ are the maximum and minimum eigenvalues of the inertia matrix $\bm{J}$ respectively. Choosing $\lambda$ satisfying Eq.\ref{Lambda} provides $\dot{V} \leq 0$ which proves asymptotic convergence of $\|\bm{\Omega}\|$ and $(1 - \beta_0)$ to zero for the control input in Eq.\ref{ControlInput}. We can immediately conclude that $\bm{\beta_v}$ goes to zero since the quaternion is unit norm constrained. The norm of the augmented state can be lower bounded as 
\begin{equation}
\|\bm{\omega}\| - \rvert 2g^2\hat{k}\rvert \|\bm{\beta_v}\| \leq \|\Omega\|
\end{equation}
The asymptotic convergence of $\bm{\beta_v}$ and $\|\bm{\Omega}\|$ to zero then lets us conclude convergence of $\lim_{t\rightarrow \infty}\|\bm{\omega}\| = 0$. We have thus shown that the vector part of the quaternion and the angular velocity go to zero asymptotically.

As demonstrated by the Lyapunov-like analysis above, the proposed control input drives the system states to zero and stabilizes the attitude of the spacecraft, while potentially avoiding the need for the high control torques from the previous literature. This will be further investigated through the simulations in the following section.
\section{Simulation Results}

The proposed controller was tested in simulations for typical values of inertia matrices for micro-satellites. Two inertia matrices $\bm{J_1}$ and $\bm{J_2}$ were used, with $\bm{J_2}$ being relatively larger:
\begin{equation}
\bm{J_1} = 
\begin{bmatrix}
3.05 & 0.14 &0.05 \\
0.14 & 2.66 & 0.12 \\
0.05 & 0.12 & 2.18
\end{bmatrix}
\end{equation} 

\begin{equation}
\bm{J_2} = 
\begin{bmatrix}
20 & 1.2 & 0.9 \\
1.2 & 17 & 1.4 \\
0.9 & 1.4 & 15
\end{bmatrix}
\end{equation}

The thruster schedule used is similar to the one used in [\citenum{JGCDSrikant}] and is given in 
Table. \ref{schedule}

\begin{table}[h!]
\centering
\begin{tabular}{||c c||}
\hline
\hline
Thruster on, s & Thruster off, s \\
\hline
0 & 3.21 \\
8.01 & 11.21 \\
16.01 & 27.21 \\
32.01 & 35.21 \\
40.01 & 43.21 \\
48.01 & 51.21 \\
\hline
\hline
\end{tabular}
\caption{Thruster on-off schedule}
\label{schedule}
\end{table}

The values of the thruster rise and fall times were chosen for these simulations to be $t_r = 10ms$. As can be seen from the table above, the thruster on period $t_{on} = 3.2 s$ and the period $T = 8 s$. Using this periodic schedule of thruster on-off times, a $\mathbb{C}^1$ function $g(t)$ was selected as shown in Fig. \ref{gt}. To avoid the steep slopes that are involved in rectangular shaped functions, $g(t)$ for this paper was chosen to be a Hermite-cubic interpolating polynomial which has the values $g(tr) = 0$, $g(t_r + \frac{t_{on}}{4}) = 1$, $g(t_r + \frac{3t_{on}}{4}) = 1$, $g(2t_r + t_{on}) = 0$ and $g(T) = 0$. This approximation of the control gain was used in all the following simulations and comparisons.

\begin{figure}[!ht]
	\centering
		\begin{subfigure}[b]{0.48\linewidth}
		\includegraphics[width=\linewidth]{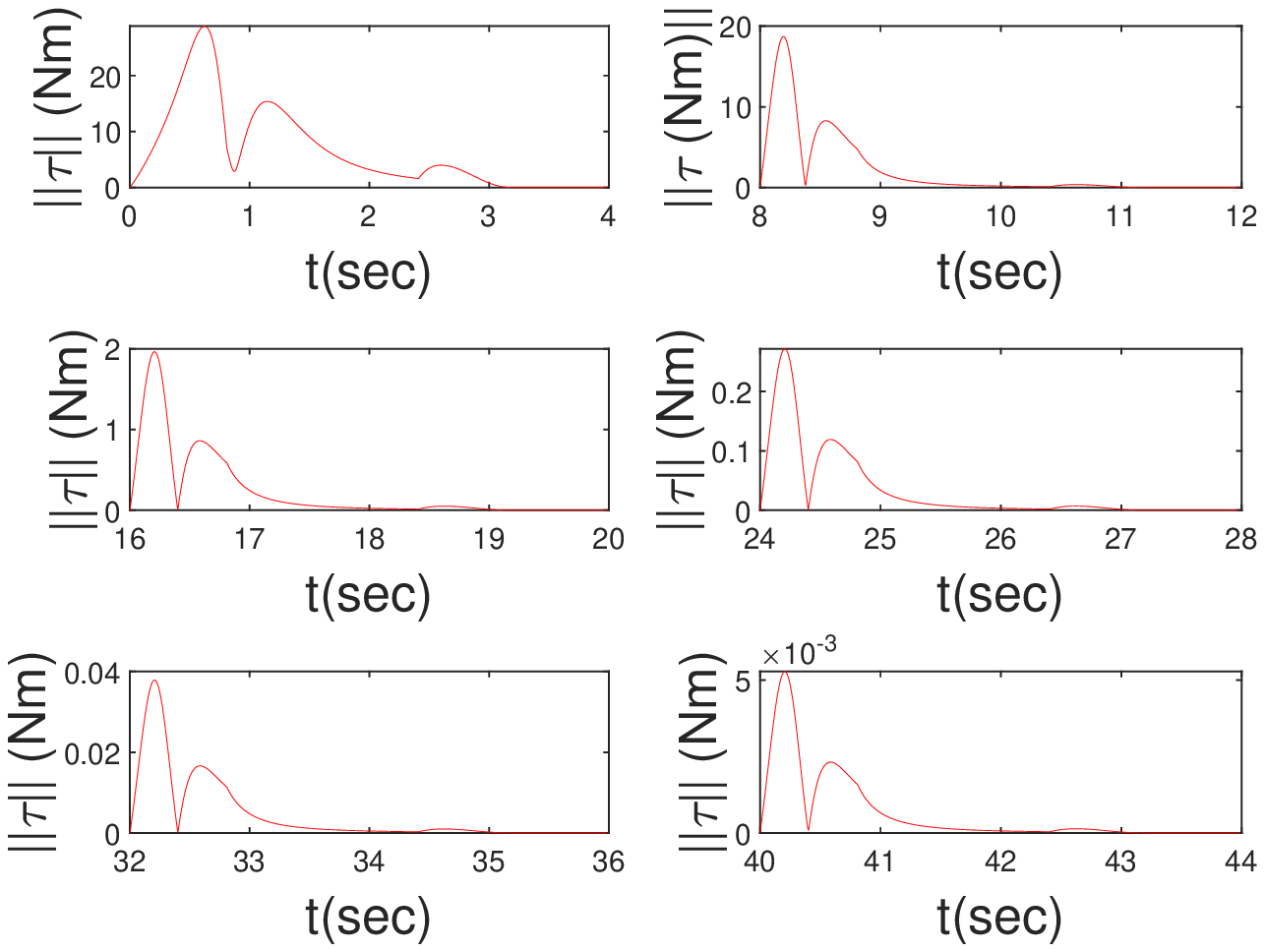}
		\subcaption{Ref. [\citenum{JGCDSrikant}]}
		\label{existingTau}
		\end{subfigure}
		\hfill
		\begin{subfigure}[b]{0.48\linewidth}
		\includegraphics[width=\linewidth]{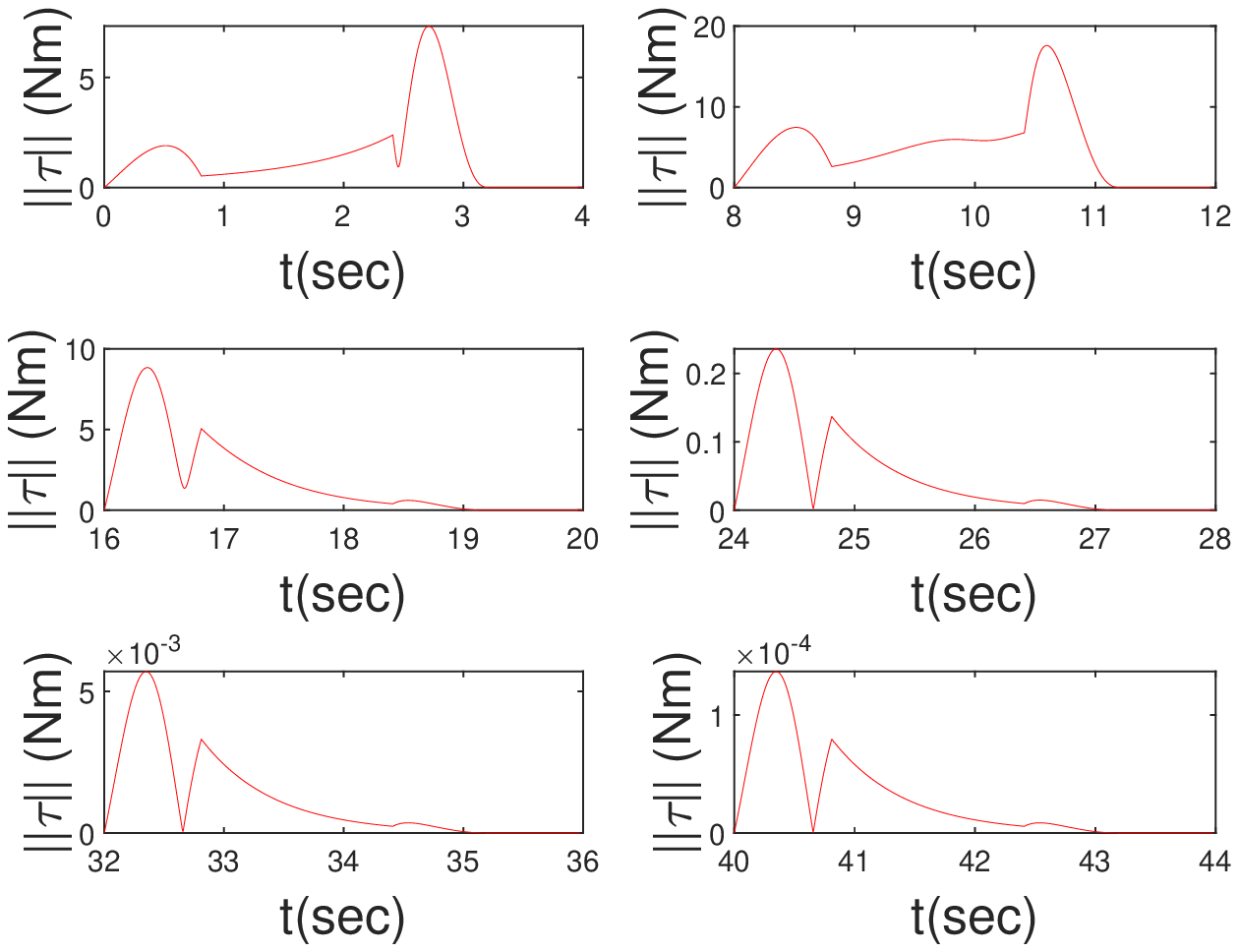}
		\subcaption{Proposed controller}
		\label{proposedTau}
		\end{subfigure}
		\vskip\baselineskip
	
		\caption{Torque Comparison for inertia $J_1$}
		
	\label{tau1}
\end{figure}

\begin{figure}[!ht]
	\centering
		\begin{subfigure}[b]{0.48\linewidth}
		\includegraphics[width=\linewidth]{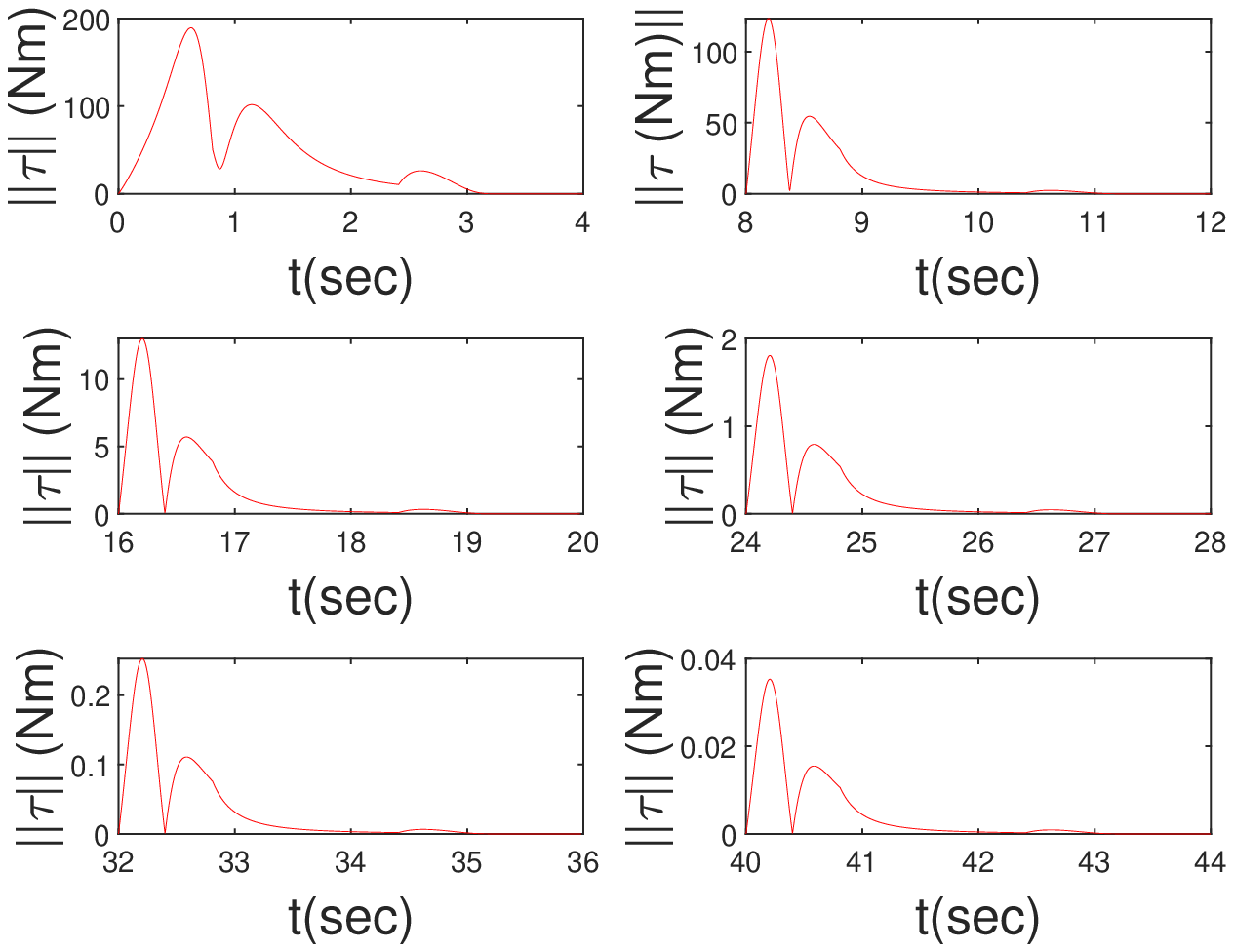}
		\subcaption{Ref. [\citenum{JGCDSrikant}]}
		\label{existingTau2}
		\end{subfigure}
		\hfill		
		\begin{subfigure}[b]{0.48\linewidth}
		\includegraphics[width=\linewidth]{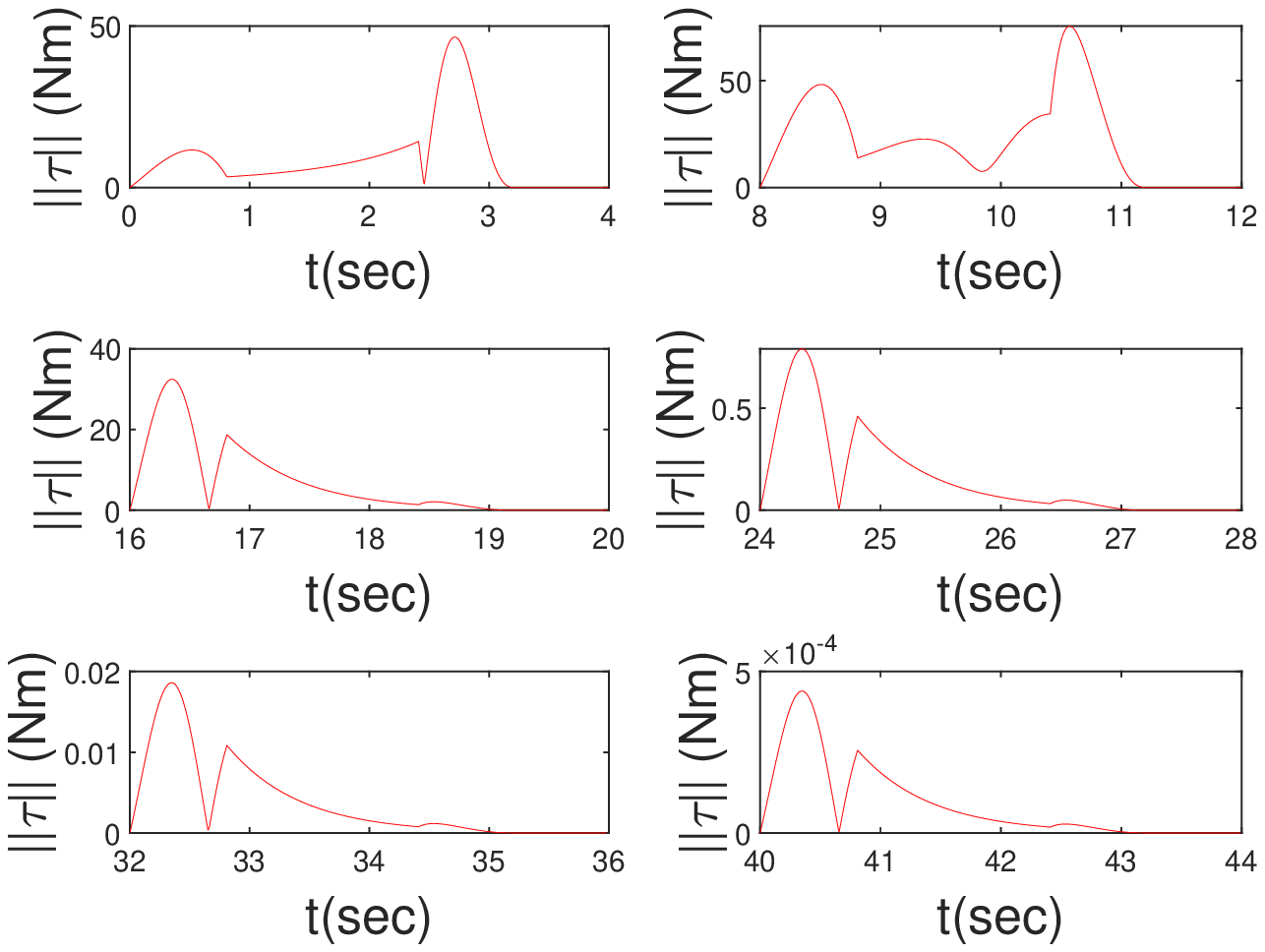}
		\subcaption{Proposed controller}
		\label{proposedTau2}
		\end{subfigure}
		\vskip\baselineskip
	
		\caption{Torque Comparison for inertia $J_2$}
		
	\label{tau2}
\end{figure}

\begin{figure}[!ht]
	\centering
		\begin{subfigure}[b]{0.48\linewidth}
		\includegraphics[width=\linewidth]{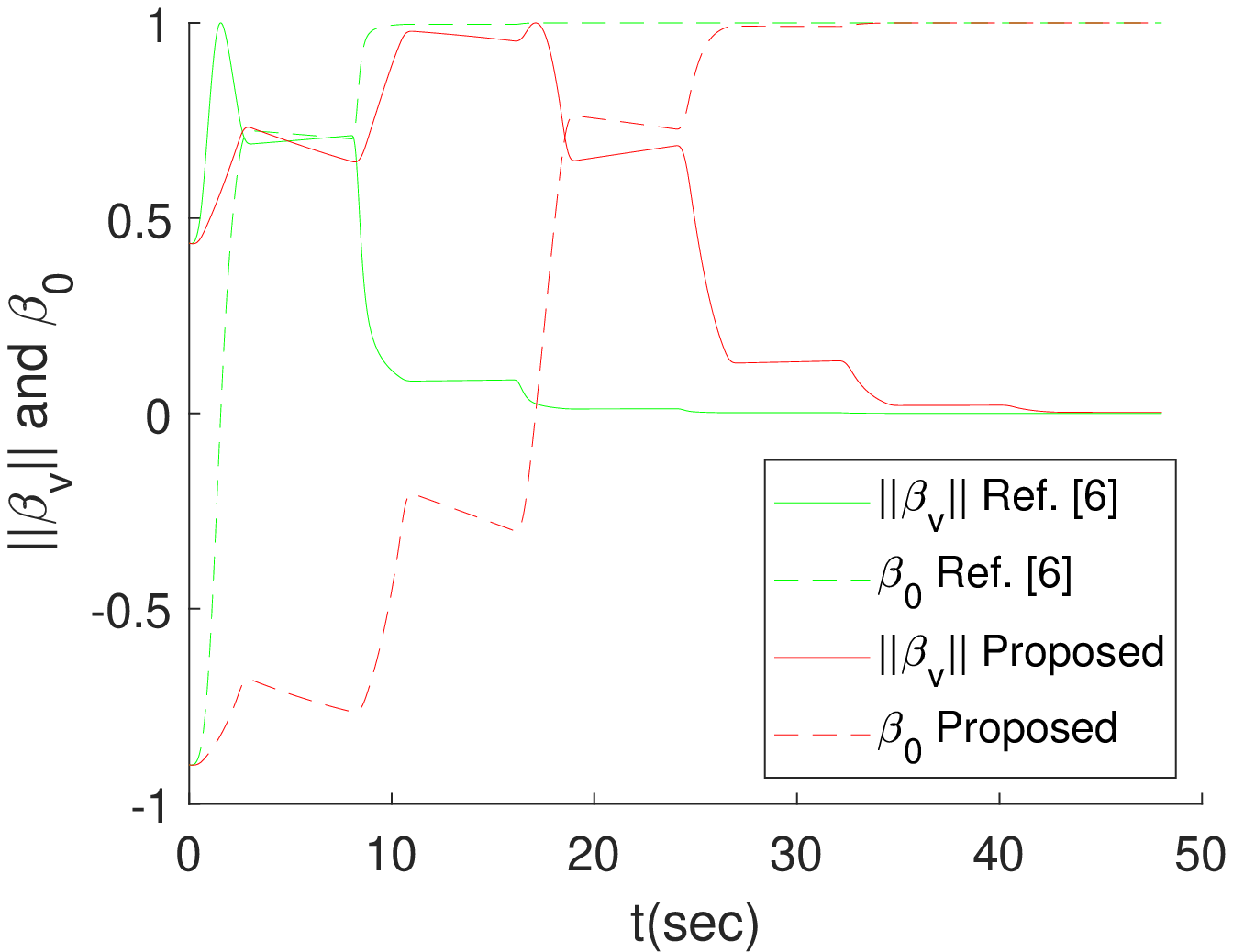}
		\subcaption{Quaternion state}
		\label{sComparison}
		\end{subfigure}
		\hfill
		\begin{subfigure}[b]{0.48\linewidth}
		\includegraphics[width=\linewidth]{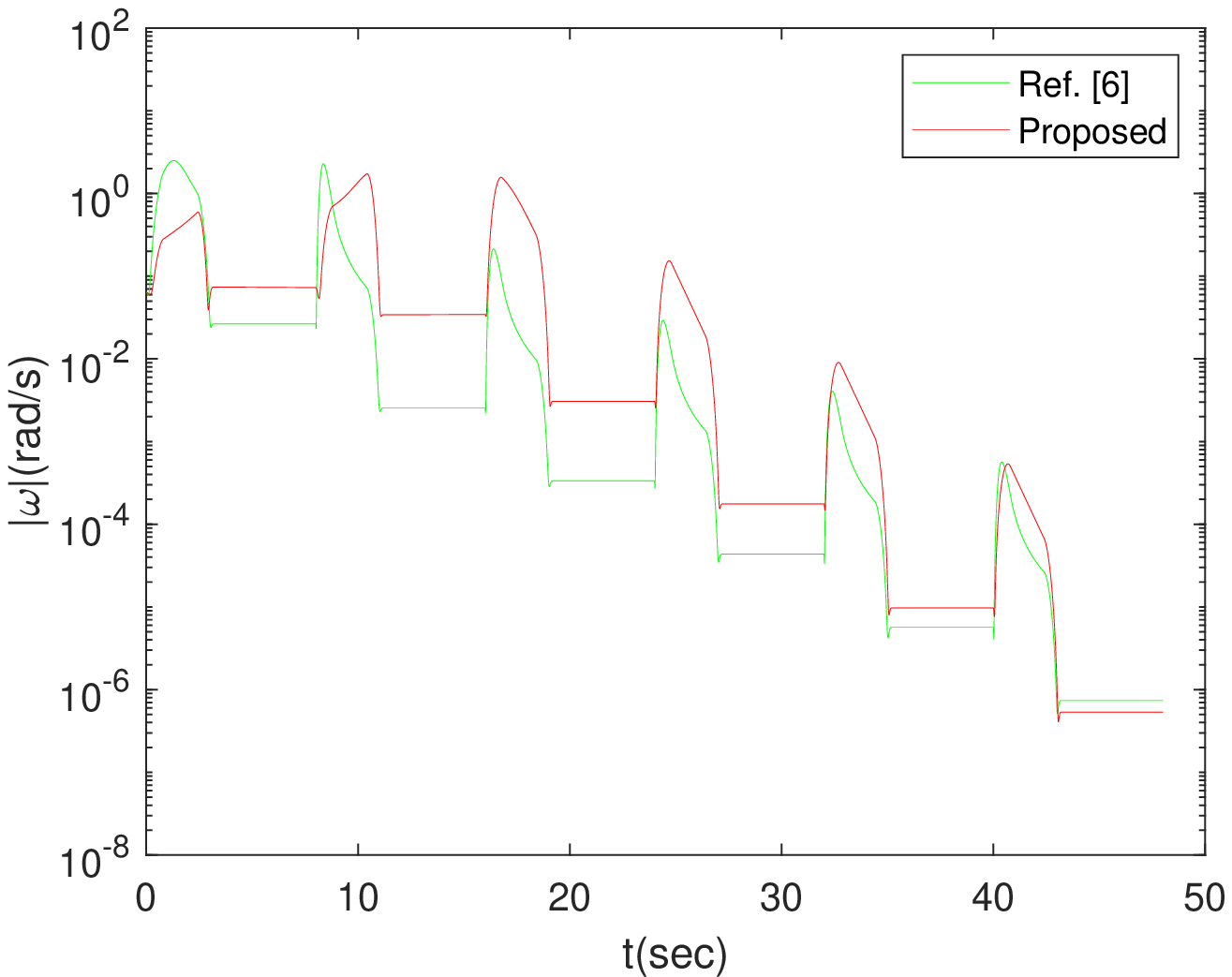}
		\subcaption{Norm of Angular Velocity}
		\label{wComparison}
		\end{subfigure}
		\vskip\baselineskip
	
		\caption{Comparison of state evolution for [\citenum{JGCDSrikant}] and proposed approach}
		
	\label{stateComparison}
\end{figure}

Fig. \ref{tau1} and Fig. \ref{tau2} show the commanded torques by the controller in [\citenum{JGCDSrikant}] and the proposed controller for the inertia matrices $\bm{J_1}$ and $\bm{J_2}$ respectively. As can be seen in both these figures, the proposed controller manages to stabilize the system with much lower spikes in the torque values in the transient but has higher values of torque compared to [\citenum{JGCDSrikant}] at later instants of time. This further demonstrates the characteristic of the proposed controller to spread out the controlled torque over a longer time period while having much lower peaks. Fig. \ref{stateComparison} shows the evolution of the attitude and angular rate states for the two controllers, for inertia $\bm{J_1}$ and further supports the observations made above about lower torques at the cost of slightly slower response. It is to be noted here that the value of the learning rate $\gamma = 0.1$ was chosen for these plots, but this rate can be chosen to be higher if quicker performance is desired, in which case the initial peaks of the commanded torques would also go up.

\begin{figure}[!ht]
\centering
	\includegraphics[width=0.8\linewidth]{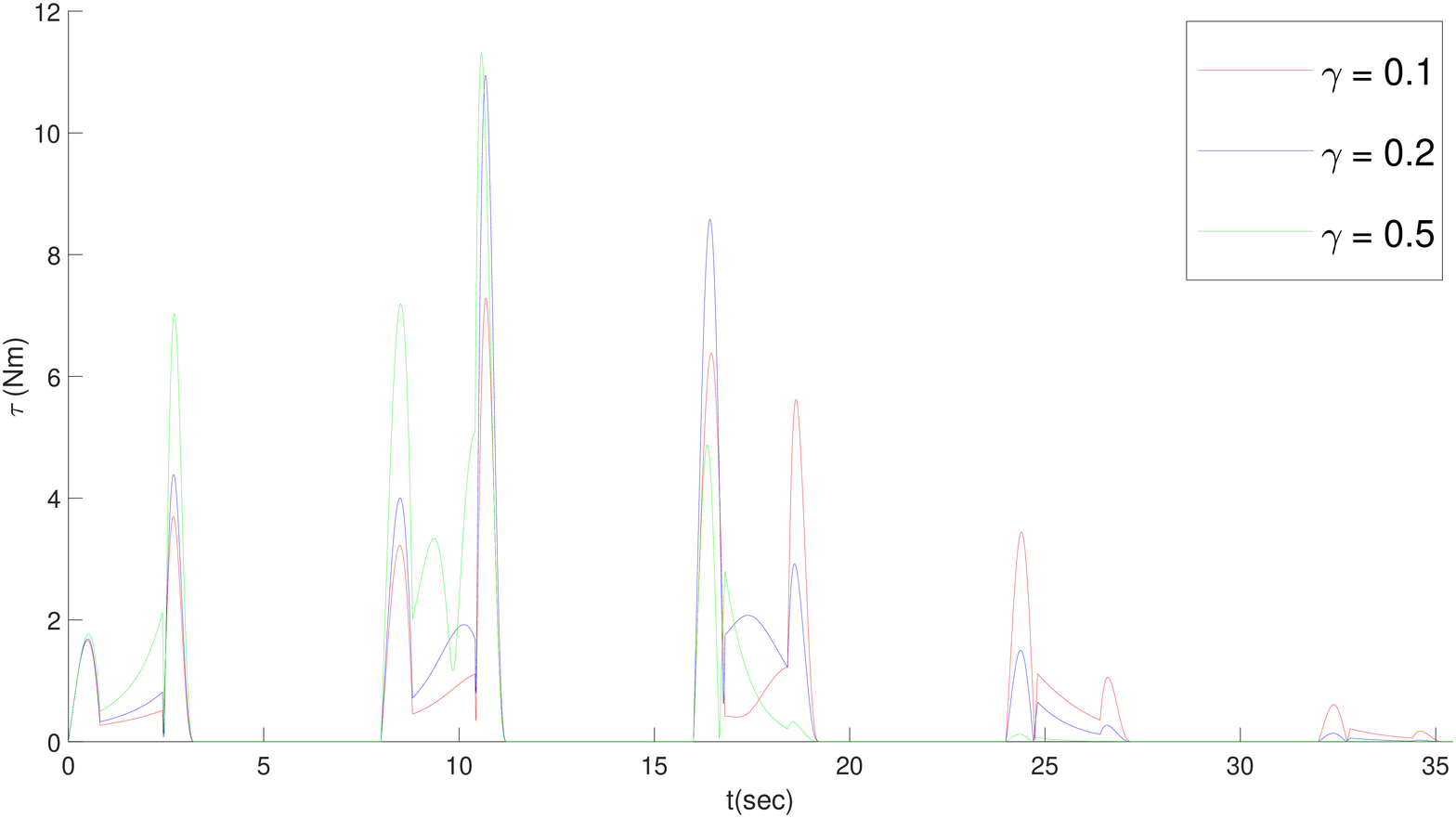}
\caption{Torque values for varying $\gamma$}
\label{gammaTau}
\end{figure}

\begin{figure}[!ht]
	\centering
		\begin{subfigure}[b]{0.48\linewidth}
		\includegraphics[width=\linewidth]{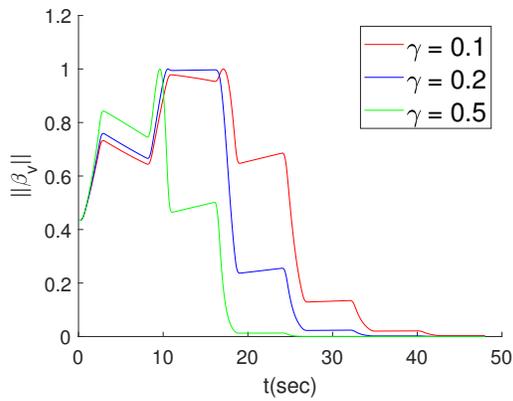}
		\subcaption{Quaternion state}
		\label{gammaS}
		\end{subfigure}
		\hfill
		\begin{subfigure}[b]{0.48\linewidth}
		\includegraphics[width=\linewidth]{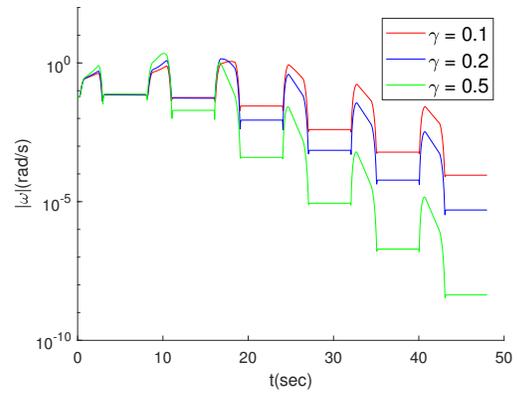}
		\subcaption{Norm of Angular Velocity}
		\label{gammaW}
		\end{subfigure}
		\vskip\baselineskip
		
	\begin{subfigure}[b]{0.48\linewidth}
		\includegraphics[width=\linewidth]{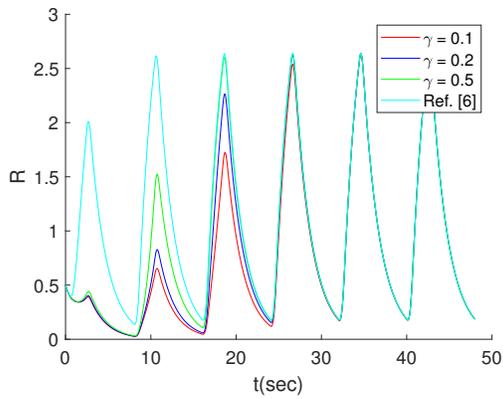}
		\subcaption{Persistence Filter state $R$}
		\label{gammaR}
		\end{subfigure}
		\hfill
		\begin{subfigure}[b]{0.48\linewidth}
		\includegraphics[width=\linewidth]{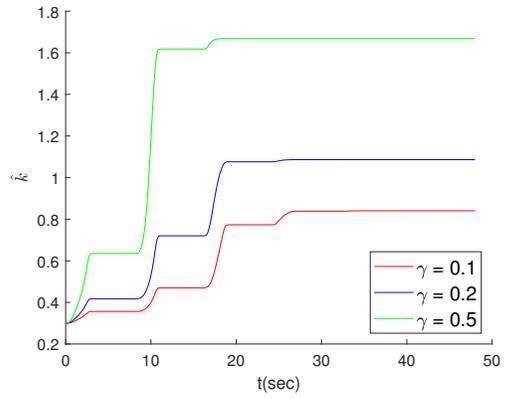}
		\subcaption{Dynamic gain $\hat{k}$}
		\label{gammaKhat}
		\end{subfigure}
		\caption{State evolution for varying $\gamma$}
		
	\label{gammaState}
\end{figure}

Fig. \ref{gammaTau} illustrates the impact of the learning parameter $\gamma$ on the torque commands where the $\gamma = 0.1$ plot starts with shorter peaks but continues to command some torque even after $t=24s$ while the $\gamma=0.5$ plot has already stabilized the system and does not need to thrust the system anymore. Fig. \ref{gammaState} and Fig. \ref{gammaW} provide the same information about the effect on the attitude and rate states, while Fig. \ref{gammaKhat} shows how the dynamic gain $\hat{k}$ goes up quicker as expected with higher $\gamma$. It is also interesting to note that the output of the persistence filter state $R$ shown in Fig. \ref{gammaR} increases differently initially for the different values of $\gamma$ as well as for the controller in [\citenum{JGCDSrikant}], but eventually converges to the same value for all the cases. This is due to the fact that the filter is dependent on the scalar part of the quaternion and the state evolves differently for each of these cases, with [\citenum{JGCDSrikant}] being the quickest due to the high gain torques and low values of $\gamma$ being the slowest. However, as $\beta_0$ settles to unity, all the controllers reach a common value of the filter output.

As is demonstrated by the simulations described above, the proposed controller is effective at stabilizing the attitude of the satellite system, while have lower spikes in the control torque and the added ability to regulate the peaks using the learning rate parameter.

\section{Conclusions}
A novel control approach was provided for the attitude stabilization of spacecraft with time varying control gains by modifying the existing persistence filter solution. The proposed controller can be readily tuned to command lower values of torque inputs in the transient and avoid the spikes seen in previous approaches, while guaranteeing asymptotic stability of the quaternion and angular rate system states. The efficacy of the controller was demonstrated in simulations and compared against existing controllers.

\bibliographystyle{AAS_publication}   
\bibliography{references}   

\end{document}